\documentclass[twocolumn,preprintnumbers, superscriptaddress, reprint, longbibliography, pra]{revtex4-1}
\usepackage{amsmath}
\usepackage{stmaryrd}
\usepackage{txfonts}
\usepackage{amssymb}
\usepackage{mathrsfs}
\usepackage{graphicx}
\usepackage{dcolumn}
\usepackage{bm}
\usepackage{braket}
\usepackage{epsfig}
\usepackage{color}
\usepackage{ulem}
\usepackage{bibunits}

\usepackage[colorlinks=true, linkcolor=blue, citecolor=blue, urlcolor=blue]{hyperref}
\begin{document}
\preprint{\href{https://doi.org/10.1103/PhysRevB.98.235116}{Y. Su and S.-Z. Lin, Phys. Rev. B {\bf 98}, 235116 (2018).}}

\title{Nontrivial topology and localization in the double exchange model with possible applications to perovskite manganites}
\author{Ying Su}
\email{yingsu@lanl.gov}
\author{Shi-Zeng Lin}
\email{szl@lanl.gov}
\affiliation{Theoretical Division, T-4 and CNLS, Los Alamos National Laboratory, Los Alamos, New Mexico 87545, USA}

\begin{abstract}

The double exchange model describing the coupling between conduction electrons and localized magnetic moments is relevant for a large family of physical systems including manganites. Here we reveal that the one dimensional double exchange model with an incommensurate magnetic elliptical spiral is a topological insulator with a Chern number $2\mathbb{Z}$ in the two dimensional space with one physical dimension and one ancillary dimension spanned by the Goldstone mode of the spiral. Moreover, the electronic states can be localized for a strong local exchange coupling. The topological protected edge states are responsible for the pumping of electron charge, and give rise to multiferroic response. Our  work uncovers hitherto undiscovered nontrivial topology and Anderson localization in the double exchange model with possible applications to perovskite manganites. 

\end{abstract}

\date{\today}
\maketitle

\section{INTRODUCTION}
Perovskite manganites have attracted considerable attention in the last decades since the discovery of the colossal magnetoresistance. The coupling between charge and spin, and other degrees of freedom results in many interesting phases with rich physical properties. It is found experimentally that in a class of compounds, such as $\mathrm{RMnO}_3$ (R: the rare earth element), the systems stabilize an incommensurate magnetic order. \cite{dagotto_colossal_2001,tokura_multiferroics_2014,dong_multiferroic_2015} This incommensurate magnetic order gives rise to a strong coupling between ferroelectricity and magnetism. \cite{PhysRevLett.95.057205,PhysRevLett.96.067601,khomskii_classifying_2009} The physics of manganites can be understood qualitatively in term of the double exchange model (DEM)  \cite{PhysRevB.73.094434}, which describes the interaction between conduction electrons and localized magnetic moments. 

The DEM is described by the Hamiltonian
\begin{equation}
\mathcal{H} =-t\sum_{\langle i,j \rangle} c_{i}^\dagger c_{j} - J \sum_i c_{i}^\dagger \mathbf{S}_i\cdot \bm{\sigma} c_{i},
\label{H}
\end{equation}
where $c_i=(c_{i,\uparrow}, c_{i,\downarrow})^\top$ is the annihilation operator of two-component spinor at the $i$-th site, $\bm{\sigma}$ is the vector of Pauli matrices in the spin space of conduction electrons, and $\mathbf{S}_i$ are the localized moments. In manganites, $\mathbf{S}_i$ describes the spins of the localized $t_{2g}$ electron, and $c_i$ is the itinerant $e_g$ electron. The first term is the nearest neighbor hopping of conduction electrons and the last term describes the local exchange coupling between the conduction electrons and localized magnetic moments, {which can be much stronger than the hopping amplitude in manganites, $J\gg t$ \cite{PhysRevLett.104.017201}}. Here we consider the limit $|S_i|\gg 1$ and treat the localized moments as classical degrees of freedom. The ground state magnetic configuration is determined by minimizing the free energy functional $\mathcal{F}(\mathbf{S}_i)$. Self-consistent calculations have shown that Eq. \eqref{H} with $\mathcal{F}(\mathbf{S}_i)$ relevant for manganites can stabilize a magnetic spiral \cite{PhysRevB.73.094434,PhysRevB.78.155121,PhysRevLett.118.027203,PhysRevX.4.031045}. {It is shown that Eq. \eqref{H} alone can stabilize incommensurate magnetic spiral for certain region of electron filling and $J$. \cite{yunoki_phase_1998}} Besides the relevance to manganites, model \eqref{H} can also describe the complex magnetic orderings in rare earth magnets due to the Ruderman-Kittel-Kasuya-Yosida interaction, heavy Fermion behavior due to Kondo singlet formation \cite{gulacsi_one_2004,HewsonBook}, and anomalous or topological Hall effect \cite{RevModPhys.82.1539}.

\begin{figure}[b]
  \begin{center}
  \includegraphics[width=8 cm]{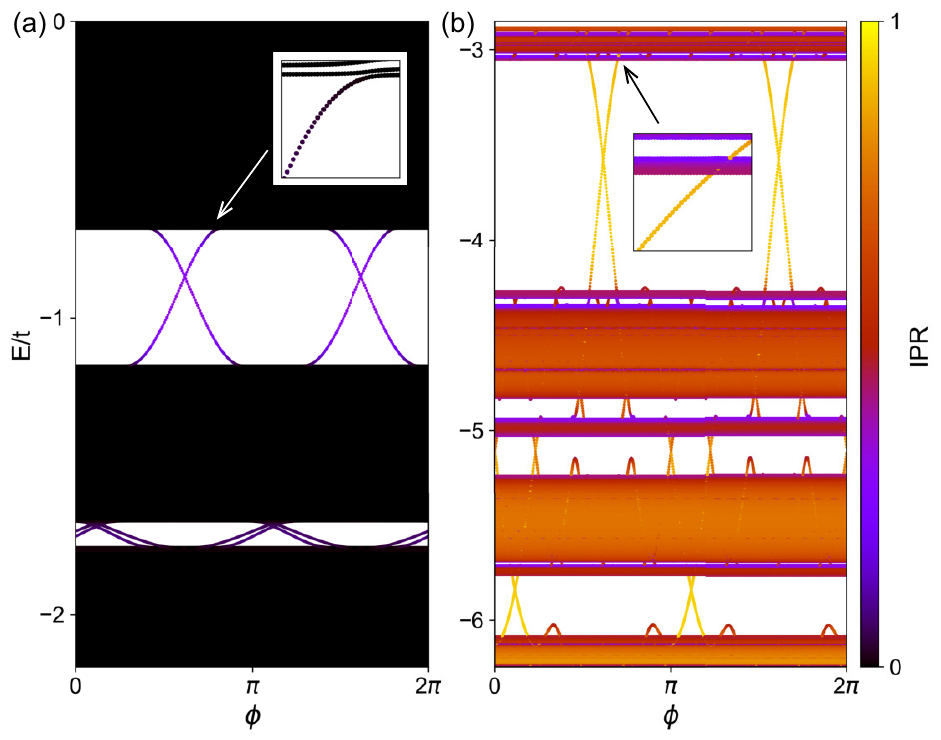}
  \end{center}
\caption{Spectra of the DEM as a function of $\phi$ for $J=t$ (a) and for $J=6t$ (b). Because the whole spectrum is symmetric with zero energy, only the negative branches are shown here. The open boundary condition is used to manifest the edge states.  The color of spectra denotes the IPR of eigenstates. The insets show the enlarged spectrum of the areas specified by the arrows. Here $\lambda=(1+\sqrt{5})/2$, {$Q=2\pi/\lambda$}, $L=610$, and $b=0.5$.} 
  \label{fig1}
\end{figure}

\begin{figure*}
  \begin{center}
  \includegraphics[width=17 cm]{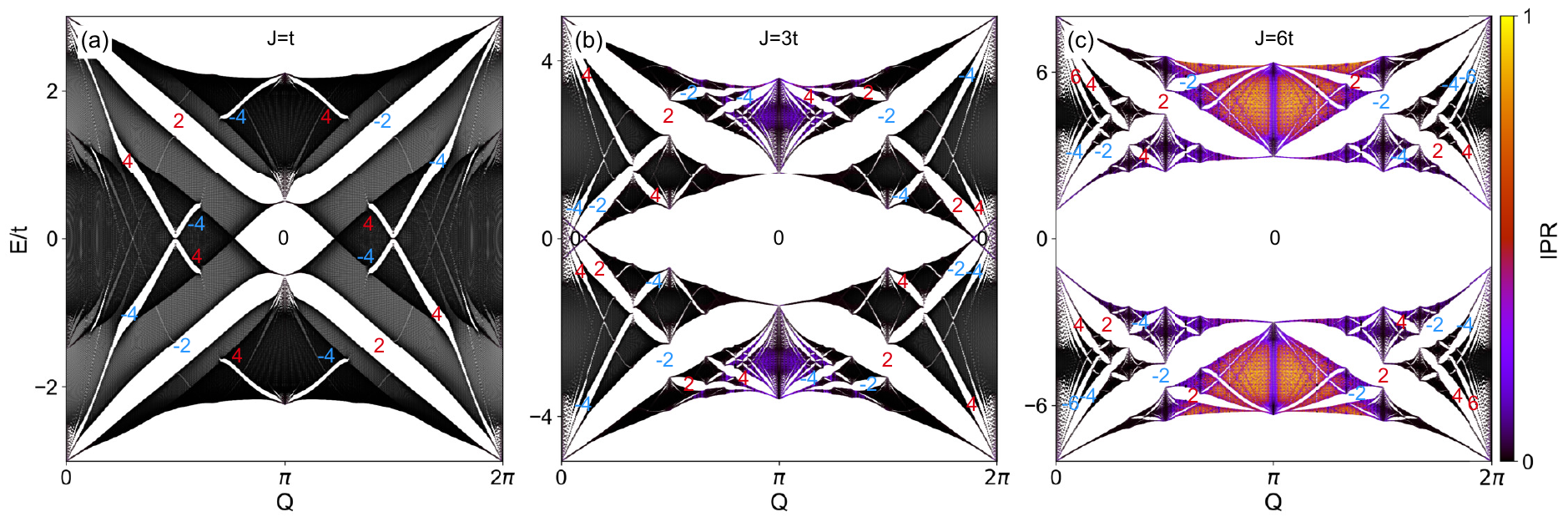}
  \end{center}
\caption{(a)-(c) Spectra of the DEM with $J=t$, $3t$, and $6t$, respectively, and as a function of $Q$. 
The periodic boundary condition is used. The numbers in the gaps are the Chern numbers. Here $L=610$, $b=0.5$, and $\phi=0$ are fixed. } 
  \label{fig2}
\end{figure*}

In this work, we study the effect of an incommensurate magnetic order on the conduction electrons. The incommensurate structure allows for the existence of a Goldstone phason mode $\phi \in [0,\ 2\pi]$, which corresponds to the spatial translation of the magnetic structure. We reveal that the one dimensional (1D) DEM is a topological Chern insulator in the manifold spanned by the real space and an ancillary dimension spanned by the Goldstone mode $\phi$. As a consequence of the bulk-edge correspondence, there exist edge states in the insulating gap. We further show that when $|\mathbf{S}_i|$ is nonuniform in space due to spin anisotropy and fluctuations, the bulk electronic states become localized. Because of the in-gap edge state, the system can pump integer electrons from one edge to the opposite edge by exciting the phason mode $\phi$, giving rise to the multiferroic response. Thus  our  work reveals hitherto undiscovered nontrivial topology and Anderson localization in the simple 1D DEM with possible applications to perovskite manganites.

\section{Model}

We start with an elliptical spiral magnetic texture 
\begin{equation}
\mathbf{S}_i(\phi) = (0,\ b \sin(Qi + \phi),\ \cos(Qi+\phi)),
\label{S}
\end{equation} 
where the lattice constant is set to unity and $b\in[0,1]$. We have also studied a more realistic magnetic configuration by minimizing $\mathcal{F}(\mathbf{S}_i)$ for a perovskite manganite as shown in Sec. \ref{sec3}, where the behavior is qualitatively the same as that for Eq. \eqref{S}. The spin texture is coplanar, therefore the berry phase of electrons induced by a non-coplanar spin texture is absent. \cite{RevModPhys.82.1539} When $b=0$ for an Ising density wave, Eq. \eqref{H} reduces to the well known Aubry-Andr\'{e}-Harper (AAH) model \cite{Aubry_Analyticity_1980,Harper_Single_1955} both in the spin up and spin down sectors, which can be mapped exactly to the two dimensional (2D) Hofstadter model \cite{HofstadterModel}. The Hofstadter model describes the 2D integer quantum Hall effect on a lattice. \cite{PhysRevB.23.5632,PhysRevLett.49.405,PhysRevLett.71.3697}  When $b=1$ for a circular spiral, one can eliminate the off-diagonal local exchange coupling in the electron spin space by performing the unitary transformation 
$\bar{c}_j =e^{-i\sigma_x(Qj+\phi)/2} c_j $, such that the local quantization axis of the conduction electron is aligned with $\mathbf{S}_i$. 
The Hamiltonian becomes 
$\mathcal{H} =-t\sum_{\langle i,j\rangle}\bar{c}_i^\dagger \left(\cos\frac{Q}{2}\sigma_0 {-i\sin \frac{Q}{2}\sigma_x}\right) \bar{c}_j- J \sum_i \bar{c}_{i}^\dagger \sigma_z \bar{c}_{i}$, which has the same translational symmetry as the lattice. Hence, no localization happens when $b=1$. When $J/t\gg 1$ or $Q\ll 1$, the off-diagonal hopping can be neglected, and $\mathcal{H}$ is reduced to the simple tight-binding model with trivial topology. Here $\sigma_0$ is the $2\times2$ identity matrix. 
As will be discussed below, non-trivial topology and Anderson localization occur in the region $0\le b < 1$.

\section{Topological phase} 
Because $\phi$ is a phason mode associated with translation of the magnetic structure, one generally would expect the electronic spectrum is invariant with respect to $\phi$. This is indeed the case for the bulk states as shown in Fig. \ref{fig1}. However, there is one notable feature i.e. the appearance of localized edge states in the gap. The degree of localization can be quantified by defining the inverse participation ratio (IPR) of the eigenstates ${\rm IPR}(E_n) ={\sum_i |\psi_n(i)|^4}/{\left(\sum_i |\psi_n(i)|^2\right)^2}$, where $E_n$ and $\psi_n(i)$ are the eigenvalue and wavefunction of the $n$-th eigenstate. \cite{PhysRevB.96.085119} It is finite for an localized state but vanishes as $1/L$ for an extended state. The existence of the edge states implies that the DEM is topological nontrivial.

In the DEM, the time-reversal symmetry is broken and the band topology can be characterized by a Chern number. The commensurate DEM shares the same band topology as the incommensurate DEM if the band gap does not close when $Q$ is varied adiabatically. As depicted in Fig. \ref{fig2} for electronic spectrum as a function of $Q$, the infinitesimal change of $Q$ does not close the gap. This allows us to approximate the irrational period $\lambda=2\pi/Q$ by a close rational number in the calculation of the Chern number \cite{Kraus_Topological_2012}. For the golden ratio $\lambda=(1+\sqrt{5})/2$, we can approximate it by $\lambda\approx F_n/F_{n-1}$, where $F_n=F_{n-1}+F_{n-2}$ is the $n$-th Fibonacci number. In the limit $L\rightarrow\infty$, the commensurate DEM is described by $\mathcal{H}(\phi) = \sum_{k=0}^{2\pi}h(k,\phi)$, where 
\begin{equation}\label{eqh}
h=-t\sum_{i=1}^{F_n} \left( e^{ik/F_n }c_{i+1}^\dagger c_{i} + {\rm H.c.}  \right) - J \sum_{i=1}^{F_n} c_{i}^\dagger \mathbf{S}_i\cdot \bm{\sigma} c_{i},
\end{equation}
Because $h(k,\phi)$ is periodic in both $k$ and $\phi$ which form a compact 2D manifold, the Chern number can be calculated as \cite{PhysRevLett.51.51}
\begin{equation}
C=\frac{1}{2\pi i} \int_0^{2\pi} dk \int_0^{2\pi} d\phi{\rm Tr}\left( \mathcal{U}\left[\partial_k \mathcal{U},\partial_\phi \mathcal{U}	\right] \right),
\label{C}
\end{equation}
where $\mathcal{U}(k, \phi ) = \sum_{E_n<E_F}\ket{\psi_n(k,\phi)}\bra{\psi_n(k,\phi)}$ is the projection operator of occupied states. Here $E_n(k,\phi)$ and $\ket{\psi_n(k,\phi)}$ are the $n$-th eigenvalue and eigenstate of $h(k,\phi)$. {The method to compute $C$ is outlined in Appendix \ref{app1}.} The Chern number of the DEM when the Fermi energy is in the gaps is denoted by an integer in Fig. \ref{fig2}. Indeed, DEM is a topological Chern insulator with the number of edge states being in agreement with the Chern number.

\section{Symmetry analysis}
\begin{figure}
  \begin{center}
  \includegraphics[width=8 cm]{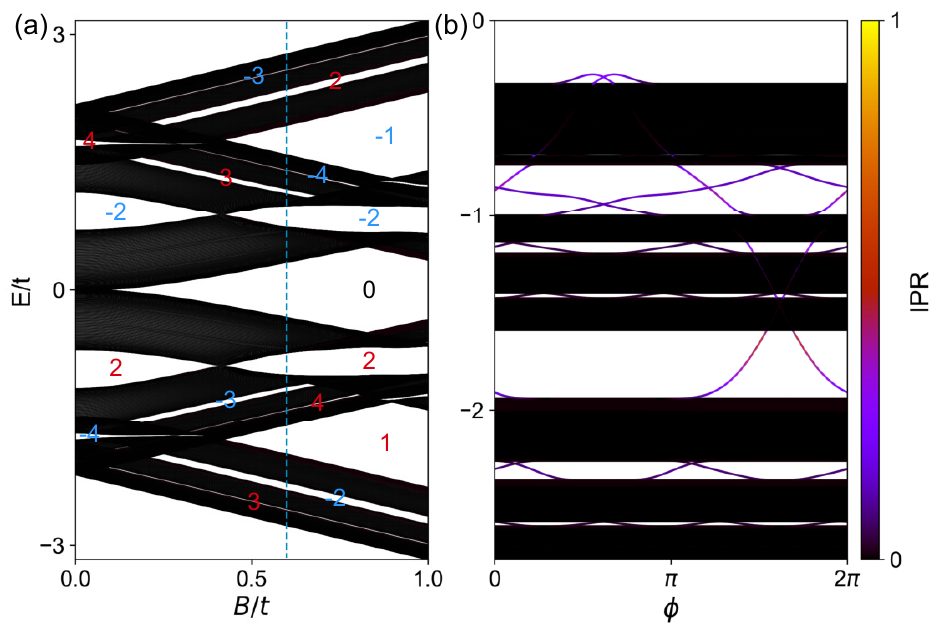}
  \end{center}
\caption{(a) Spectrum of the perturbed DEM in Eq. (\ref{Zeeman}) as a function of $B$ and with $J=t$, $b=0.5$, and $\phi=0$.
Here we take $F_{15}=610$ and $F_{14}=377$ to approximate the golden ratio $\lambda\approx F_{15}/F_{14}$ and $Q=2\pi/\lambda$. The error of the approximated golden ratio is less than $2\times 10^{-6}$ and does not change the topology.
The numbers mark the Chern numbers for the gaps. (b) Spectrum for $B=0.6t$ [marked by the blue dashed line in (a)] as a function of $\phi$. Periodic boundary condition is used in (a), while open boundary condition is used in (b).} 
  \label{fig3}
\end{figure}

There are four important features of the energy spectrum of DEM. (1) The spectrum is invariant when $\phi$ is shifted by $\pi$. (2) All the Chern numbers are even. (3) The spectrum is symmetric with respect to zero energy and $Q=\pi$ {[hence only the negative part of the whole spectrum is shown in Figs. \ref{fig1} and \ref{fig3}(b)]}. (4) The Chern number changes sign with respect to $E=0$ and $Q=\pi$. As a consequence, the insulating state around $E=0$ and $Q=\pi$ corresponding to antiferromagnetic arrange of $\mathbf{S}_i$ is topological trivial. 

The properties (1) and (2) are originated from the spin degree of freedom of the system. When $b=0$, Eq. \eqref{H} describes two copies of Hofstadter model in the up and down spin channels. The up and down spin sectors are related by $\pi$ shift of $\phi$. For $0\le b<1$, Eq. \eqref{H} is invariant under global rotation of electron spins and localized moments. This means that rotation of electron spins is equivalent to the rotation of local moments in the opposite direction. Thus the rotation of electron spins around $x$ direction by $\pi$ yields $\mathcal{R}_x(\pi)\mathcal{H}(Q, \phi)\mathcal{R}_x^{-1}(\pi)=\mathcal{H}(Q, \phi+\pi)$. Therefore, the edge states must appear in pairs, i.e. if there is an edge sate at $\phi$, then there must be another edge state at $\phi+\pi$ with electron spin rotated along the $x$ direction by $\pi$. Because of the bulk-edge correspondence, the Chern number of the gapped states must be even, and the Hamiltonian \eqref{H} is characterized by a $2\mathbb{Z}$ Chern number. 

The properties (3) and (4) are due to the chiral symmetry and another rotation operation of Eq. \eqref{H}. Under the global rotation of the electron spins along the $z$ axis by $\pi$,  Eq. \eqref{H} is transformed into $\mathcal{R}_z(\pi)\mathcal{H}(Q, \phi)\mathcal{R}_z^{-1}(\pi)=\mathcal{H}(-Q, -\phi)$, where the rotation of $\bm{S}_i$ in the opposite direction corresponds to reverse sign of $Q$ and $\phi$. Therefore, the spectrum is symmetric under the transformation $Q\rightarrow-Q$. Because the unitary transformation does not change the band topology, and $H(-Q,\phi)$ and $H(-Q,-\phi)$ have the opposite Chern numbers according to Eq. \eqref{C}, $C$ is odd under the transformation $Q\rightarrow -Q$. 

In addition, Eq. \eqref{H} has the chiral symmetry, defined by the transformation  $\mathcal{O}_{\rm C} c_{i,\sigma} \mathcal{O}_{\rm C}^{-1} = (-1)^i c_{i,\bar{\sigma}}$ that flips the electron spin, and add a minus sign to the electron operators at one sub-lattice. Here $\mathcal{O}_{\rm C}$ is the chiral operator that is unitary, $\mathcal{O}_{\rm C}^2 = 1$. Under the chiral transformation, the Hamiltonian changes sign, $\mathcal{O}_{\rm C}\mathcal{H}(\phi)\mathcal{O}_{\rm C}^{-1} = -\mathcal{H}(\phi)$. Hence the presence of chiral symmetry guarantees a symmetric energy spectrum with respect to zero energy. Therefore the Chern number for all the states $\ket{\psi}$ above $E>E_g$ and below $E<-E_g$ is the same, $C(E>E_g)=C(E<-E_g)$. Here $E_g>0$ is an arbitrary energy that lies in an insulating gap. Because the Chern number for all states of a Hamiltonian vanishes, $C(E>E_g)+C(E<E_g)=0$, this immediately means that the Chern number for the occupied states with $E<E_g$ and $E<-E_g$ has the opposite sign, $C(E<E_g)=-C(E<-E_g)$.

The $2\mathbb{Z}$ topological classification can be reduced to $\mathbb{Z}$ if the $\pi$-spin rotation symmetry is broken. We consider the perturbation by an external magnetic field along the $z$ direction
\begin{equation}
\mathcal{H}'(\phi,B)=\mathcal{H}(\phi) - B \sum_{i} c_{i}^\dagger \sigma_{z} c_{i}.
\label{Zeeman}
\end{equation}
Here we assume the Zeeman coupling of $\mathbf{S}_i$ is much weaker compared to other interactions in $\mathcal{F}(\mathbf{S}_i)$ and neglect this coupling in $\mathcal{F}(\mathbf{S}_i)$. The Zeeman field breaks the $\pi$-spin rotation symmetry but respects the chiral symmetry. As displayed in Fig. \ref{fig3}, the energy gaps close and reopen as $B$ increases, indicating topological phase transitions. The Chern numbers and the number of the edge states can be both even and odd, see Figs. \ref{fig3}(a) and \ref{fig3}(b). The spectrum is no longer invariant under the $\pi$ shift of $\phi$.

\section{Localization transition}
\begin{figure}
  \begin{center}
  \includegraphics[width=8 cm]{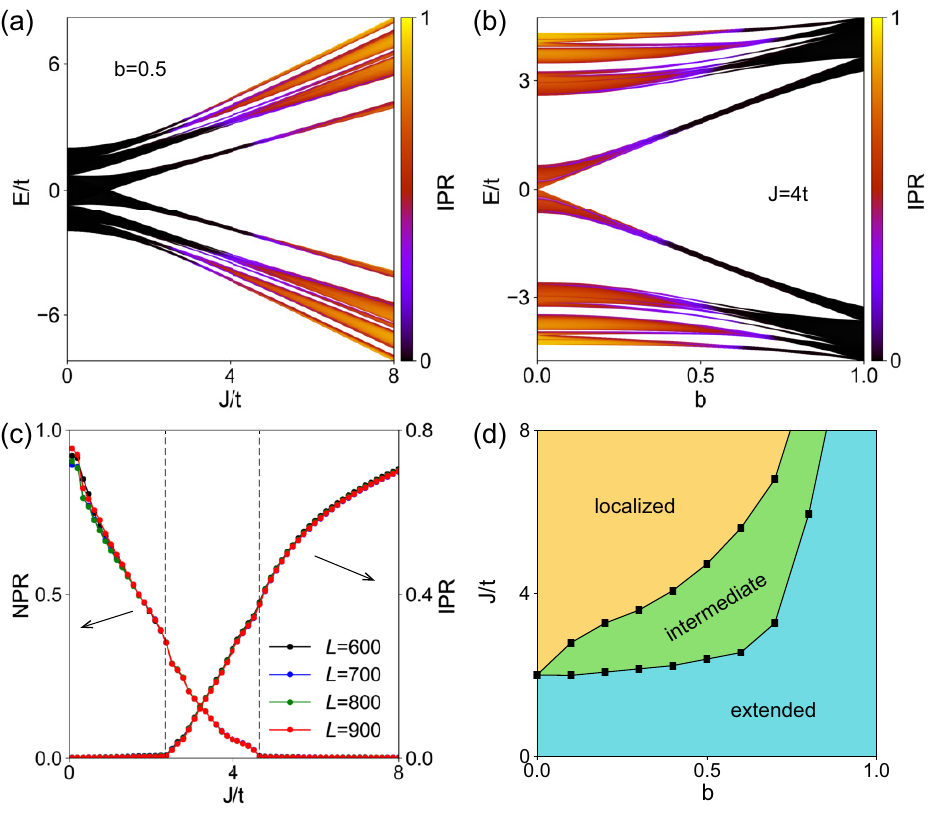}
  \end{center}
\caption{(a) Spectrum of the DEM as a function of $J$ for $b=0.5$. (b) Spectrum of the DEM as a function of $b$ for $J=4t$. Here $Q=2\pi/\lambda$, $\phi=0$, and $L=610$ are fixed for both cases.
(c) Averaged NPR and IPR as a function of $J$ for different lengths and with $b=0.5$. The two dashed lines enclose the intermediate region. (d) The phase diagram in the $b$-$J$ plane. The blue, green, yellow regions are respectively the extended, intermediate, localized phases. 
} 
  \label{fig4}
\end{figure}

The incommensurate and nonuniform $|\mathbf{S}_i|$ generates an incommensurate local potential on electrons, which can lead to the localization of electron wave functions. For the AAH
model when $b=0$, all the electronic states are extended when $J<2t$ but are localized at $J>2t$. For an elliptical spiral, both the edge and bulk states are localized at a large $J$, as indicated by a large IPR in Fig. \ref{fig1}(b). Meanwhile, the band with spin antiparallel to $\mathbf{S}_i$, $E>0$, and band with spin parallel to $\mathbf{S}_i$, $E<0$, are well separated. As shown in Fig. \ref{fig4}(a), when $J$ increases, the system evolves from the extended phase to the localized phase featured by the increase of IPR. Different from the AAH model, the critical $J$ depends on $E$ when $0<b<1$. Namely, there is a mobility edge, and the extended and localized states coexist in the transition region. To determine the coexistence region, we introduce the normalized participation ratio (NPR), ${\rm NPR}(E_n) = {1}/{L\sum_i |\psi_n(i)|^4}$, which is the counterpart of IPR and approaches zero for localized states. As displayed in Fig. \ref{fig4}(c), there exists an region with coexisting localized and extended states, where both the NPR and IPR averaged over all eigenstates are nonzero. The phase diagram of localized and extended states is presented in Fig. \ref{fig4} (d). When $b$ is close to $1$, the spatial modulation of the incommensurate potential becomes weaker and it becomes more difficult to localize the electron states, as displayed in Fig. \ref{fig4} (b). 

Similar to the integer quantum Hall systems, the Chern number of the DEM can still be defined even though all states are localized at a large $J$. \cite{Kraus_Topological_2012,PhysRevB.35.2188} As shown in Fig. \ref{fig2} (c), topological insulator with $2\mathbb{Z}$ Chern number and with localized bulk states can be achieved. Hence the DEM is a realization of the topological Anderson insulator \cite{PhysRevLett.102.136806,PhysRevLett.103.196805,PhysRevB.93.214206}.

\section{Elliptical spiral in perovskite manganites}\label{sec3}
Here we study the electronic spectrum with a more realistic spin structure obtained by minimizing the free energy functional $\mathcal{F}(\mathbf{S})$. We consider magnetic spiral in perovskite manganites as an example \cite{PhysRevB.73.094434,PhysRevB.78.155121} and the analysis is valid for spirals stabilized by frustrated or Ruderman-Kittel-Kasuya-Yosida interaction. We will show that the thermal effect and spin anisotropy can distort a circular spiral into an elliptical spiral, and the resulting spectrum is qualitatively similar to that obtained by using the simple ansatz in Eq. \eqref{S}. 

\begin{figure}[t]
  \begin{center}
  \includegraphics[width=8 cm]{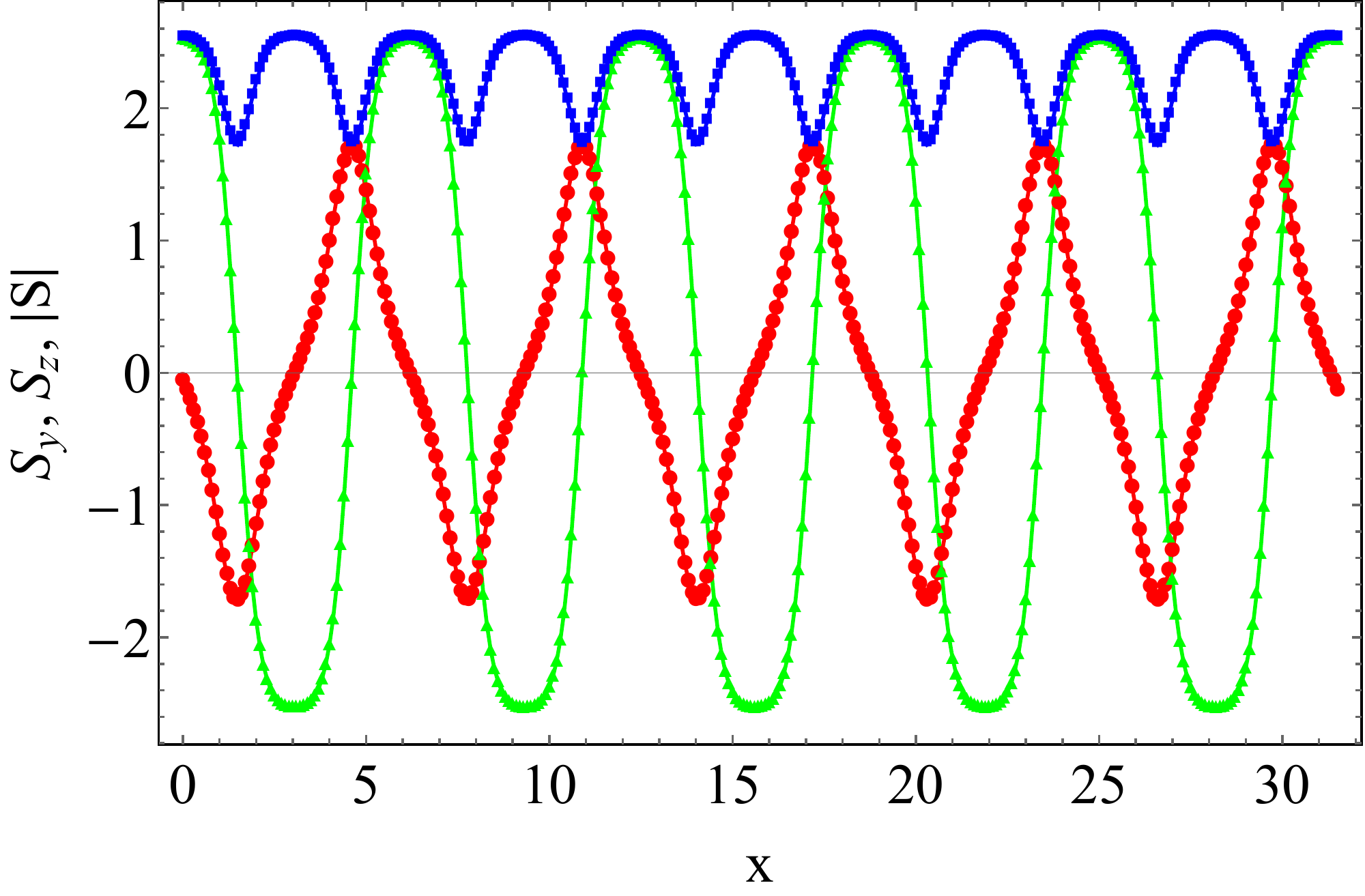}
  \end{center}
\caption{Spin configuration obtained by minimizing $\mathcal{F}$ in Eq. \eqref{Fs1}. The red, green and blue curves are for $S_y$, $S_z$ and $|\mathbf{S}|$ respectively. Here $A=5.0$.} 
  \label{fs1}
\end{figure}

\begin{figure}[b]
  \begin{center}
  \includegraphics[width=8 cm]{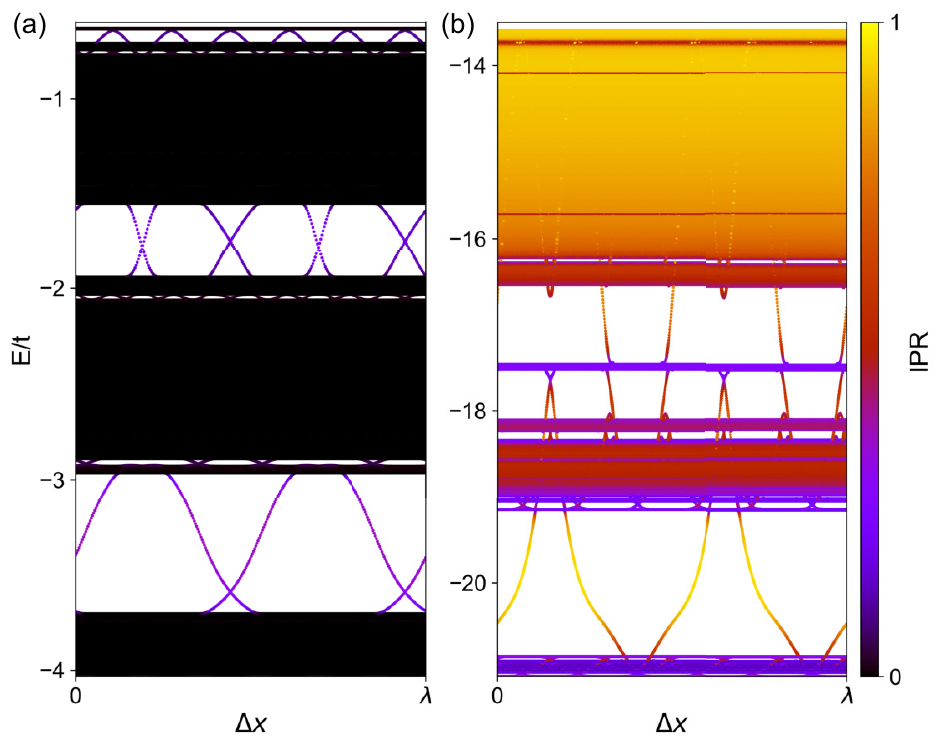}
  \end{center}
\caption{Spectra of the DEM with the spin configuration shown in Fig. \ref{fs1}. (a) for $J=t$ and (b) for $J=8t$. Here $\Delta x$ is the shift of the spin configuration with respect to the tight-binding lattice, $\lambda$ is the period of the spin configuration, and $L=600$.}
  \label{fs2}
\end{figure}

We consider the dimensionless Ginzburg-Landau free energy functional \cite{bak_theory_1980}
\begin{align}\label{Fs1}
{\cal F} = \int dx\left[ { - \frac{1}{2}{{\left( {\bf{S}} \right)}^2} + \frac{1}{4}{{\left( {{{\bf{S}}^2}} \right)}^2} + \frac{1}{2}{{\left( {\nabla {\bf{S}}} \right)}^2} + {\bf{S}}\cdot\left( {\nabla  \times {\bf{S}}} \right) - \frac{A}{2}S_z^2} \right],
\end{align}
where the last second term is the Dzyaloshinskii-Moriya interaction, and the last term with $A>0$ is the easy axis anisotropy. At finite temperature, $|\mathbf{S}(x)|$ is allowed to fluctuate in space.  When $A=0$, the ground state configuration is a circular spiral $S_x=0$, $S_y=b \sin(Q x)$ and $S_z=b \cos(Q x)$ with $Q=-1$ and $b=\sqrt{2}$. For a nonzero $A$, we calculated the magnetic configuration by minimizing $\mathcal{F}$ numerically and the result is shown in Fig. \ref{fs1}, where an elliptical spiral with spatially modulated $|\mathbf{S}(x)|$ is stabilized.

We then calculate the electronic spectrum using the obtained $\mathbf{S}$. The lattice parameter for the tight-binding chain is $a=1$. As displayed in Fig. \ref{fs2}, the insulating gap is topological nontrivial and states becomes localized for a large $J$. Here $\Delta x$ denotes the shift of the spin structure with respect to the tight-binding lattice. The period of the spiral is $\lambda=6.29$, see Fig. \ref{fs1}.

\begin{figure}[t]
  \begin{center}
  \includegraphics[width=8 cm]{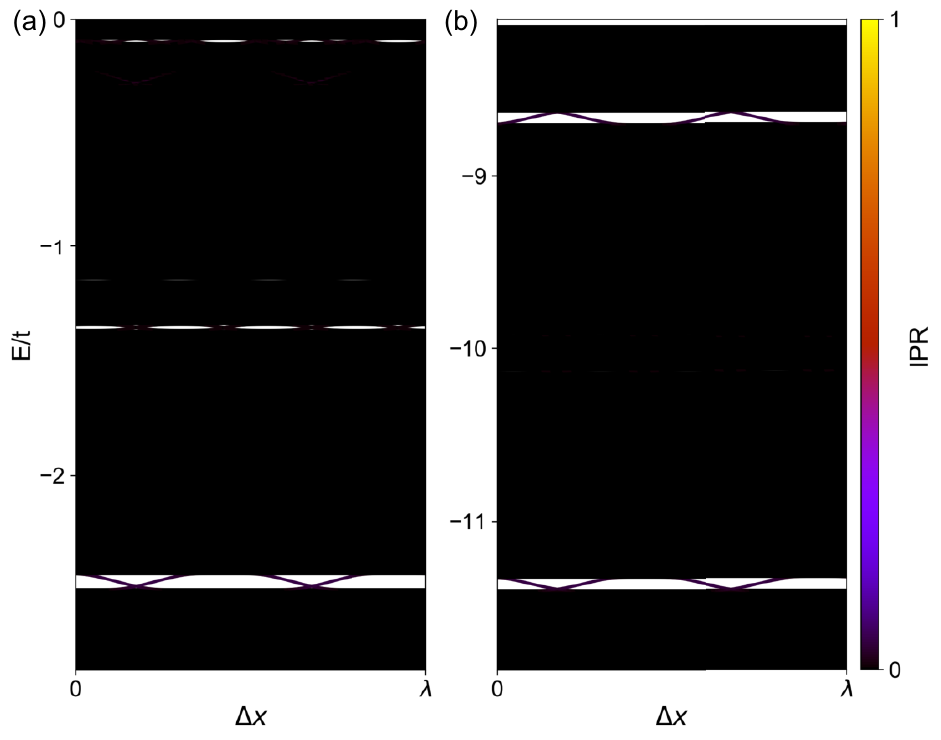}
  \end{center}
\caption{Spectra of the DEM with the spin configuration given by Eq. \eqref{cnsn} with $\gamma=0.7$. (a) for $J=t$ and (b) for $J=10t$. Here $\Delta x$ is the shift of the spin configuration with respect to the tight-binding lattice, $\lambda$ is the period of the spin configuration, and $L=600$. }
  \label{fs4}
\end{figure}

At zero temperature where $|\mathbf{S}|$ is uniform in space, the spin texture distorted by an easy axis anisotropy can be described by the ansatz
\begin{align}\label{cnsn}
\mathbf{S}(x)=(0,\ \mathrm{sn}(x, \gamma),\ \mathrm{cn}(x, \gamma)),
\end{align} 
where sn and cn are the Jacobi elliptic functions and they reduce to the sine and cosine functions when $\gamma=0$. The period of the magnetic structure is $\lambda=4K(\gamma)$ with $K(\gamma)$ being the complete elliptic integral of the first kind. The electronic spectrum as a function of $\Delta x$ is shown in Fig. \ref{fs4}, where the nontrivial topology remains. However, there is no localization of electronic states, no matter how strong the local exchange coupling is.

The absence of localization for arbitrary strong $J$ can be understood using the unitary transformation, $\bar{c}_j =e^{-i\sigma_x\theta_j/2} c_j $. The Hamiltonian becomes
\begin{equation}
\mathcal{H} =-t\sum_{\langle i,j\rangle}\bar{c}_i^\dagger \left(\cos\frac{\Delta\theta_{ij}}{2}\sigma_0-i\sin \frac{\Delta\theta_{ij}}{2}\sigma_x\right) \bar{c}_j- J \sum_i 
\bar{c}_{i}^\dagger \sigma_z \bar{c}_{i},
\end{equation} 
where $\Delta \theta_{ij}=\theta_i-\theta_j$ and $\theta_j\equiv \arctan(S_{j,z}/S_{j,y})$ with $S_{j,z}$, $S_{j,y}$ being the $z$ and $y$ components of the local magnetic moment at the tight-binding site $j$. The local exchange field becomes uniform in space, while the hopping amplitudes are modulated. The incommensurate modulation of the order of $t$ is independent of $J$ and is not enough to induce localization. 
{To reveal the topological edge states, it requires to tune the electron filling such that the Fermi energy lies within the topological energy gaps. }

\begin{figure}[b]
  \begin{center}
  \includegraphics[width=8 cm]{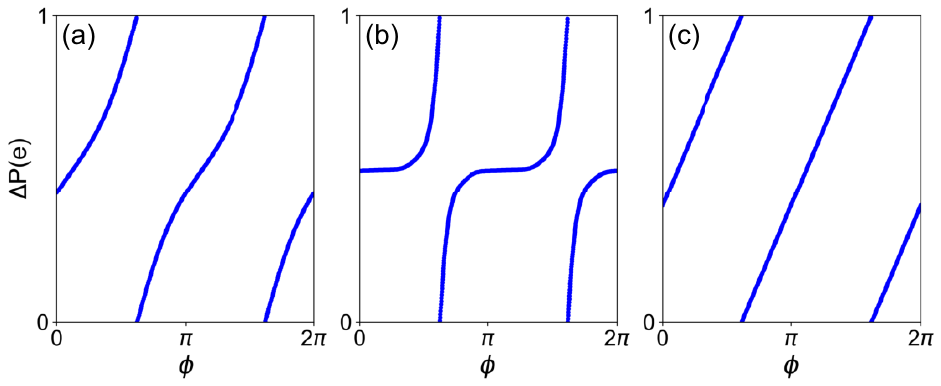}
  \end{center}
\caption{(a) and (b) Polarization from Eq. (\ref{P}) of the approximated commensurate DEM as a function of $\phi$. $J=t$ for (a) and $J=6t$ for (b) correspond to the spectra in Figs. \ref{fig1} (a) and \ref{fig1}(b), respectively, and Fermi energies $E_F=-t$ and $-3.6t$ are in the gaps where the number of edge modes is 2. (c) Polarization for $J=6t$ is also calculated using $P(\phi)=\frac{e}{L}\sum_{E_n<E_F,i}\bra{\psi_n(i)}i\ket{\psi_n(i)}$ with the open boundary condition and $L=610$.} 
  \label{fig5}
\end{figure}

\begin{figure}[t]
  \begin{center}
  \includegraphics[width=8 cm]{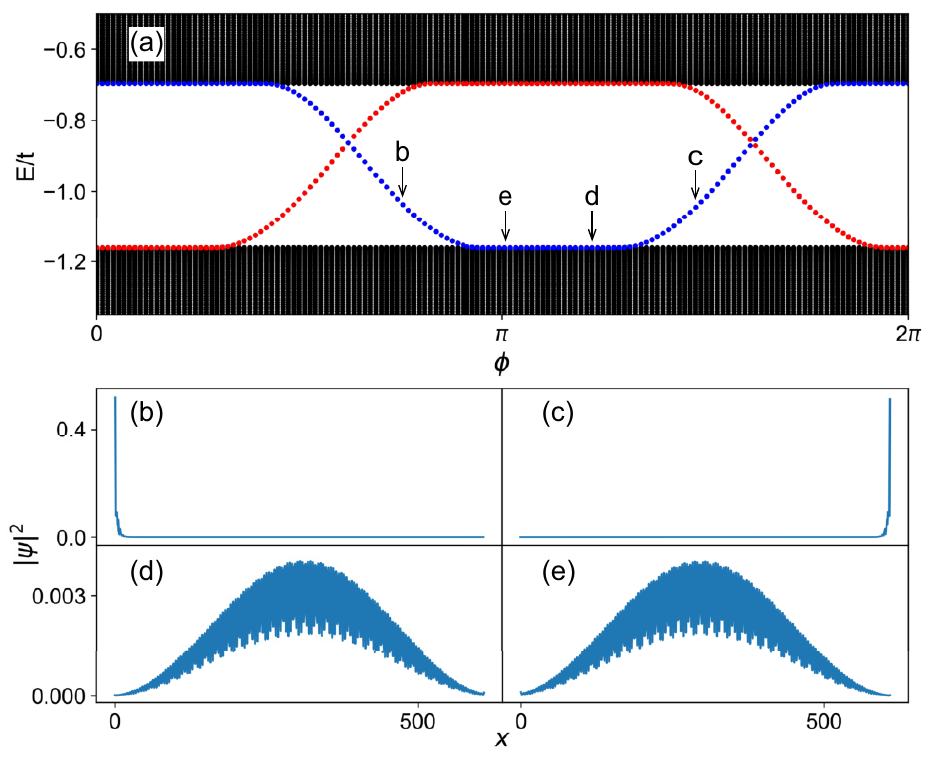}
  \end{center}
\caption{(a) Evolution of the edge states in the upper gap of Fig. \ref{fig1}(a). The corresponding wave functions are shown in (b), (c), (d) and (e). }
  \label{fs5}
\end{figure}

\section{Adiabatic charge pumping and polarization}
The presence of magnetic spiral breaks the inversion symmetry explicitly, and the insulating state can have nonzero electric polarization. For extended states, the electric polarization can be calculated through the Wannier center and is expressed in term of the Berry phase of the occupied bands \cite{RevModPhys.66.899}
\begin{equation}
P(\phi) = \frac{ie}{2\pi}\sum_{E_n<E_F}\int_0^{2\pi} dk \bra{\psi_n(k,\phi)}\partial_k \ket{\psi_n(k,\phi)},
\label{P}
\end{equation}
where $e$ is the electron charge. $P(\phi)$ is not gauge invariant, but its difference $\Delta P(\phi)$ is gauge invariant. Here we define $\Delta P(\phi)=P(\phi)-{\rm min}(P)$. When the phason mode $\phi$ is excited, for instance, by a thermal gradient, the electron is pumped from one edge to the other if the gap remains open. In the adiabatic limit, $\Delta P(\phi)$ both for extended states and localized states is shown in Fig. \ref{fig5}. It is clear that there are $2e$ charge pumped from the left to the right edge when $\phi$ advances by $2\pi$. The number of pumped charge is the same as the number of edge states and the Chern number. \cite{PhysRevB.27.6083} Because of the $\pi$-spin rotation symmetry, the edge states at $\phi$ and $\phi+\pi$ have the opposite spin polarization, therefore no net spin is pumped in a complete cycle. For the localized states, $P$ can be obtained directly using the definition $P(\phi)=\frac{e}{L}\sum_{E_n<E_F,i}\bra{\psi_n(i)}i\ket{\psi_n(i)}$, which gives the consistent description as shown in Fig. \ref{fig5} (c). The Anderson localization in the bulk cannot prevent the charge transfer between two edges through the bulk. This is completely different from the topological trivial Anderson insulators. Moreover, the Berry phase is proportional to the polarization as $\gamma(\phi)=2\pi P(\phi)/e$, see Eq. \eqref{P}. As shown in Fig. \ref{fig5}, the Berry phase winds twice when $\phi$ is swept from 0 to $2\pi$. The winding of Berry phase is the Wilson loop and the winding number equals the Chern number. \cite{PhysRevB.84.075119,TopoBook} We remark that the electric polarization in the present model has different origin as that induced by magnetic spiral.  \cite{PhysRevLett.95.057205,PhysRevLett.96.067601,khomskii_classifying_2009}

We next study the evolution of the edge states during the charge pumping from the left to the right edge. For the extended bulk states at small $J$, the states localized at edges merge into bulk bands but remain at the top or bottom of the bands as $\phi$ increases [see Fig. \ref{fig1} (a) and Sec. \ref{sec5}]. On the other hand, for the localized bulk states, the evolution of the edge states is completely different. To transport an electron from the left to the right edge, all the localized states shift towards right. Meanwhile, there are quantum tunnelings of localized electronic states at different positions in the bulk. [see Fig. \ref{fig1} (b) and Sec. \ref{sec5}].

\section{Evolution of edge states during charge pumping}\label{sec5}
In this section, we provide details about the evolution of edge states during charge pumping both for the extended and localized bulk states. For extended states as shown in Fig. \ref{fs5}, the edge states merge into the bulk states and remain at the top or bottom of the bulk bands. The electronic states localized at the edges become extended when they merge into the bulk band.

\begin{figure}[htb]
  \includegraphics[width=8 cm]{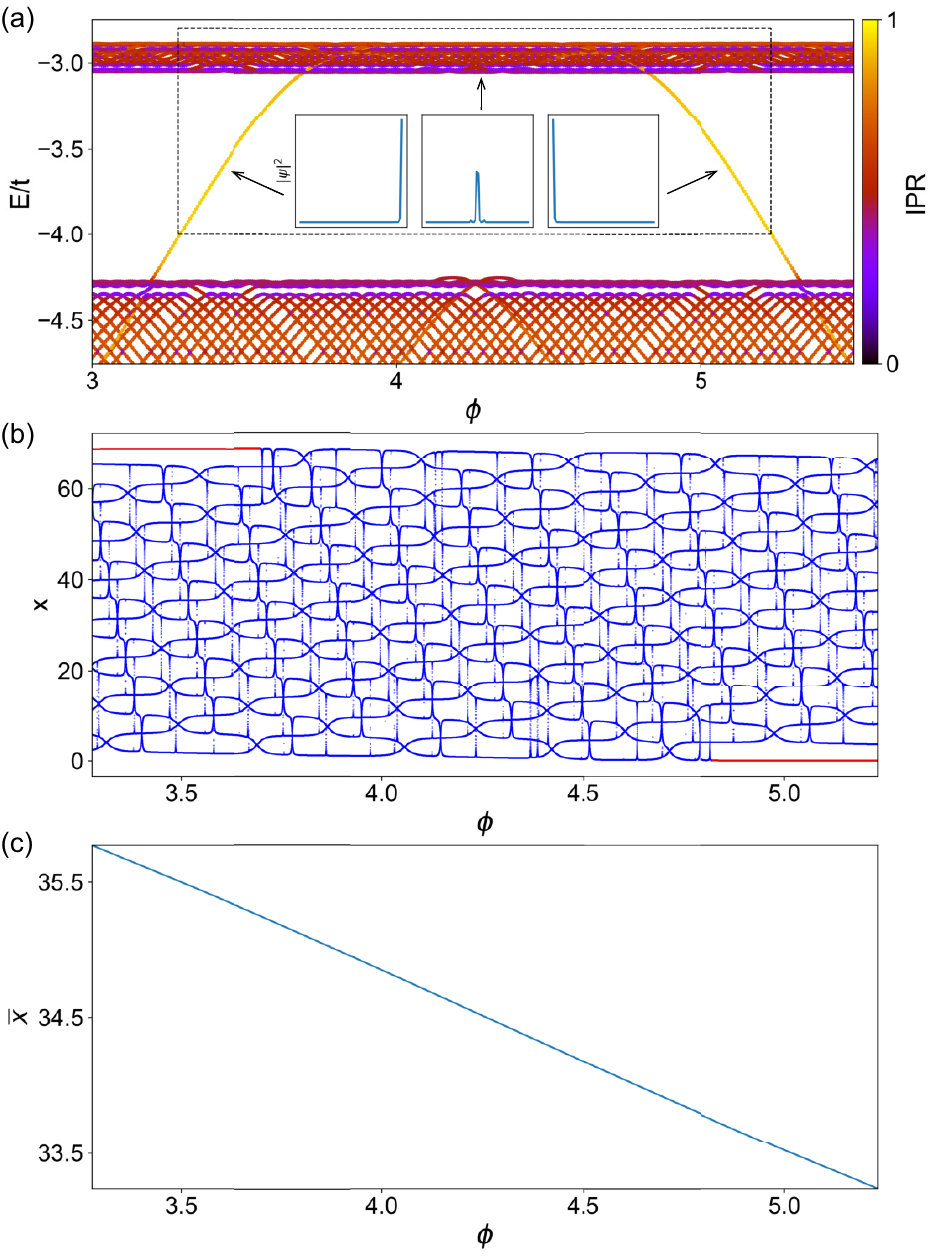}
\caption{(a) Evolution of the edge states in the upper gap of Fig. \ref{fig1}(b). For clarity, we reduce the length of chain to $L=70$. The edge states change, however, the bulk band gap and topology remain invariant. The insets show the wave functions of eigenstates specified by arrows. (b) Localization centers $x_n=\sum_i\bra{\psi_n(i)}i\ket{\psi_n(i)}$ of all the states enclosed by the dashed square in (a). The red lines mark the position of edge states. (c) The averaged localization center, $\bar{x}=\frac{1}{N}\sum_n x_n$, where $N$ is the total number of selected energy levels. }
  \label{fs6}
\end{figure}

The situation for the localized states is entirely different as shown in  Fig. \ref{fs6}. Because all the electronic states are localized, when we advance $\phi$, there are quantum tunneling of electronic states as indicted by a sharp change of $x_n$, i.e. electrons exchange positions. Here $x_n=\sum_i\langle \psi_n(i)|i|\psi_n(i)\rangle$ is the localization center of the $n$-th localized eigenstate. Meanwhile, there is an overall shift of the electron positions from the right to the left, as shown by the center of mass of all the selected electronic states $\bar{x}=\frac{1}{N}\sum_n x_n$ where $N$ is the number of selected energy levels as shown in Fig. \ref{fs6} (a).

\section{DISCUSSION}
In this work, we have unearthed the {nontrivial} topology and localization in the simple 1D DEM. The discussion on 1D DEM can be generalized to higher dimensions when the magnetic structure is only modulated in one direction (the $x$ direction). The lattice translation symmetry in the transverse direction is preserved and we can introduce the lattice momentum $k_y$ and $k_z$, and the Hamiltonian becomes
\begin{equation}
\begin{split}
\mathcal{H}_{3D} =&-t\sum_{\langle i,j \rangle,k_y,k_z} c_{i,k_y,k_z}^\dagger c_{j,k_y,k_z} - J \sum_{i,k_y,k_z} c_{i,k_y,k_z}^\dagger \mathbf{S}_i\cdot \bm{\sigma}c_{i,k_y,k_z} \\
&-2t'\sum_{i,k_y,k_z}\left(\cos k_y+\cos k_z\right)c_{i,k_y,k_z}^\dagger c_{i,k_y,k_z},
\end{split}
\end{equation}
where the last term is due to the hopping of conduction electrons in the transverse direction. $k_y$ and $k_z$ enter as a chemical potential. The energy spectrum oscillates as a function of $k_y$ and $k_z$, and the oscillation amplitude is $8t'$ in 3D, where $t'$ is the transverse hopping energy. The energy gaps of the 1D DEM survive in higher dimension as long as $t'$ is small enough, and the nontrivial topology survives. For a large $J/t$, the localization persists at higher dimensions, where the wave functions are localized (extended) in the $x$ ($y$ and $z$) direction. For an intermediate $J/t$, the mobility edge depends on $k_y$ and $k_z$ . {The incommensurate magnetic structure can be stabilized in perovskite manganites and the present study of the DEM thus implies the possible nontrivial topology and localization in perovskite manganites. }

The topology of the DEM is defined with an extra ancillary dimension associated with the Goldstone mode of the magnetic structure. The physical systems correspond to the edge of the topological Chern insulator in the higher ancillary dimensions. The ancillary dimension can be traversed by exciting the phason mode of the magnetic structure, and the topological index is equivalent to the number of pumped electrons. The topological protected edge state only exists in certain region of the phason variable, and can be measured in experiment, for instance, by scanning tunneling microscope that can detect the topological edge states by measuring the local density of states. The localization of electronic states driven by strong local exchange coupling and the incommensurability of elliptical spiral leads to enhancement of the resistivity, therefore can be accessed by transport measurement. Contrary to the conventional mechanisms of ferroelectricity generated by a magnetic texture in type-II multiferroics \cite{khomskii_classifying_2009}, the ferroelectric response in our model is caused by the nontrivial topology of electronic spectrum due to the presence of a magnetic spiral. By sweeping the phase $\phi$, electrons are pumped from one end to the other through the bulk of the chain resulting in a current. The current can be detected by transport measurement to manifest the ferroelectric response.

\section{Acknowledgements}
The authors thank Qimiao Si and Cristian Batista for helpful discussions.  Computer resources for numerical calculations were supported by the Institutional Computing Program at LANL. 
This work was carried out under the auspices of the U.S. DOE Award No. DE-AC52-06NA25396 through the LDRD program, and was supported by the Center for Nonlinear Studies at LANL.

\appendix
\section{Efficient algorithm for Chern number calculation }\label{app1}
Direct calculation of the Chern number from Eq. \eqref{C} requires to discretize the effective Brillouin zone (spanned by $k$ and $\phi$) into a lattice consists of small plaquettes of size $\Delta k\times\Delta\phi$. The convergence of Chern number depends on $1/\Delta k$ and $1/\Delta\phi$, and is slow. Therefore, the calculation can be heavy when there are many energy bands and the dimension of projection operator matrix is big. To speed up the convergence of Chern number, we employ an efficient algorithm by using the $U(1)$ link variable \cite{fukui_chern_2005}. For any two directly linked sites of the lattice, the link tensor is defined as 
\begin{equation}
\begin{split}
&T_k^{m,n}(k,\phi) = \braket{\psi_m(k,\phi)|\psi_n(k+\Delta k,\phi)}, \\
&T_\phi^{m,n}(k,\phi) = \braket{\psi_m(k,\phi)|\psi_n(k,\phi+\Delta \phi)},
\end{split}
\end{equation}
where $E_m, E_n<E_F$. For small $\Delta k$ and $\Delta \phi$, the link tensor encodes the phase difference between nearest neighbor eigenstates. The accumulation of the phase difference is given by the $U(1)$ line variable
\begin{equation}
U_\mu(k,\phi) = \frac{\text{det}[T_\mu(k,\phi)]}{|\text{det}[T_\mu(k,\phi)]|},
\end{equation}
where $\mu=k$ or $\phi$. For a closed loop along the links of a plaquette, the Berry phase is
\begin{equation}
\gamma(k,\phi)=i\ln U_k(k,\phi)U_\phi(k+\Delta k,\phi)U_k^{-1}(k,\phi+\Delta\phi)U_\phi^{-1}(k,\phi),
\end{equation} 
which is gauge-invariant.
The total Barry phase over the whole effective Brillouin zone is nothing more than the summation of Berry phases of all plaquettes and the Chern number is
\begin{equation}
C=\frac{1}{2\pi} \sum_{k,\phi} \gamma(k,\phi).
\end{equation}
In this algorithm, the Chern number converges much faster than the direct calculation of Eq. \eqref{C}.

\bibliography{references}

\begin{thebibliography}{34}%
\makeatletter
\providecommand \@ifxundefined [1]{%
 \@ifx{#1\undefined}
}%
\providecommand \@ifnum [1]{%
 \ifnum #1\expandafter \@firstoftwo
 \else \expandafter \@secondoftwo
 \fi
}%
\providecommand \@ifx [1]{%
 \ifx #1\expandafter \@firstoftwo
 \else \expandafter \@secondoftwo
 \fi
}%
\providecommand \natexlab [1]{#1}%
\providecommand \enquote  [1]{``#1''}%
\providecommand \bibnamefont  [1]{#1}%
\providecommand \bibfnamefont [1]{#1}%
\providecommand \citenamefont [1]{#1}%
\providecommand \href@noop [0]{\@secondoftwo}%
\providecommand \href [0]{\begingroup \@sanitize@url \@href}%
\providecommand \@href[1]{\@@startlink{#1}\@@href}%
\providecommand \@@href[1]{\endgroup#1\@@endlink}%
\providecommand \@sanitize@url [0]{\catcode `\\12\catcode `\$12\catcode
  `\&12\catcode `\#12\catcode `\^12\catcode `\_12\catcode `\%12\relax}%
\providecommand \@@startlink[1]{}%
\providecommand \@@endlink[0]{}%
\providecommand \url  [0]{\begingroup\@sanitize@url \@url }%
\providecommand \@url [1]{\endgroup\@href {#1}{\urlprefix }}%
\providecommand \urlprefix  [0]{URL }%
\providecommand \Eprint [0]{\href }%
\providecommand \doibase [0]{http://dx.doi.org/}%
\providecommand \selectlanguage [0]{\@gobble}%
\providecommand \bibinfo  [0]{\@secondoftwo}%
\providecommand \bibfield  [0]{\@secondoftwo}%
\providecommand \translation [1]{[#1]}%
\providecommand \BibitemOpen [0]{}%
\providecommand \bibitemStop [0]{}%
\providecommand \bibitemNoStop [0]{.\EOS\space}%
\providecommand \EOS [0]{\spacefactor3000\relax}%
\providecommand \BibitemShut  [1]{\csname bibitem#1\endcsname}%
\let\auto@bib@innerbib\@empty
\bibitem [{\citenamefont {Dagotto}\ \emph {et~al.}(2001)\citenamefont
  {Dagotto}, \citenamefont {Hotta},\ and\ \citenamefont
  {Moreo}}]{dagotto_colossal_2001}%
  \BibitemOpen
  \bibfield  {author} {\bibinfo {author} {\bibfnamefont {Elbio}\ \bibnamefont
  {Dagotto}}, \bibinfo {author} {\bibfnamefont {Takashi}\ \bibnamefont
  {Hotta}}, \ and\ \bibinfo {author} {\bibfnamefont {Adriana}\ \bibnamefont
  {Moreo}},\ }\bibfield  {title} {\enquote {\bibinfo {title} {Colossal
  magnetoresistant materials: the key role of phase separation},}\ }\href
  {\doibase 10.1016/S0370-1573(00)00121-6} {\bibfield  {journal} {\bibinfo
  {journal} {Physics Reports}\ }\textbf {\bibinfo {volume} {344}},\ \bibinfo
  {pages} {1--153} (\bibinfo {year} {2001})}\BibitemShut {NoStop}%
\bibitem [{\citenamefont {Tokura}\ \emph {et~al.}(2014)\citenamefont {Tokura},
  \citenamefont {Seki},\ and\ \citenamefont
  {Nagaosa}}]{tokura_multiferroics_2014}%
  \BibitemOpen
  \bibfield  {author} {\bibinfo {author} {\bibfnamefont {Yoshinori}\
  \bibnamefont {Tokura}}, \bibinfo {author} {\bibfnamefont {Shinichiro}\
  \bibnamefont {Seki}}, \ and\ \bibinfo {author} {\bibfnamefont {Naoto}\
  \bibnamefont {Nagaosa}},\ }\bibfield  {title} {\enquote {\bibinfo {title}
  {Multiferroics of spin origin},}\ }\href
  {http://stacks.iop.org/0034-4885/77/i=7/a=076501} {\bibfield  {journal}
  {\bibinfo  {journal} {Rep. Prog. Phys.}\ }\textbf {\bibinfo {volume} {77}},\
  \bibinfo {pages} {076501} (\bibinfo {year} {2014})}\BibitemShut {NoStop}%
\bibitem [{\citenamefont {Dong}\ \emph {et~al.}(2015)\citenamefont {Dong},
  \citenamefont {Liu}, \citenamefont {Cheong},\ and\ \citenamefont
  {Ren}}]{dong_multiferroic_2015}%
  \BibitemOpen
  \bibfield  {author} {\bibinfo {author} {\bibfnamefont {Shuai}\ \bibnamefont
  {Dong}}, \bibinfo {author} {\bibfnamefont {Jun-Ming}\ \bibnamefont {Liu}},
  \bibinfo {author} {\bibfnamefont {Sang-Wook}\ \bibnamefont {Cheong}}, \ and\
  \bibinfo {author} {\bibfnamefont {Zhifeng}\ \bibnamefont {Ren}},\ }\bibfield
  {title} {\enquote {\bibinfo {title} {Multiferroic materials and
  magnetoelectric physics: symmetry, entanglement, excitation, and topology},}\
  }\href {https://doi.org/10.1080/00018732.2015.1114338} {\bibfield  {journal}
  {\bibinfo  {journal} {Advances in Physics}\ }\textbf {\bibinfo {volume} {64}}
  (\bibinfo {year} {2015})}\BibitemShut {NoStop}%
\bibitem [{\citenamefont {Katsura}\ \emph {et~al.}(2005)\citenamefont
  {Katsura}, \citenamefont {Nagaosa},\ and\ \citenamefont
  {Balatsky}}]{PhysRevLett.95.057205}%
  \BibitemOpen
  \bibfield  {author} {\bibinfo {author} {\bibfnamefont {Hosho}\ \bibnamefont
  {Katsura}}, \bibinfo {author} {\bibfnamefont {Naoto}\ \bibnamefont
  {Nagaosa}}, \ and\ \bibinfo {author} {\bibfnamefont {Alexander~V.}\
  \bibnamefont {Balatsky}},\ }\bibfield  {title} {\enquote {\bibinfo {title}
  {Spin current and magnetoelectric effect in noncollinear magnets},}\ }\href
  {\doibase 10.1103/PhysRevLett.95.057205} {\bibfield  {journal} {\bibinfo
  {journal} {Phys. Rev. Lett.}\ }\textbf {\bibinfo {volume} {95}},\ \bibinfo
  {pages} {057205} (\bibinfo {year} {2005})}\BibitemShut {NoStop}%
\bibitem [{\citenamefont {Mostovoy}(2006)}]{PhysRevLett.96.067601}%
  \BibitemOpen
  \bibfield  {author} {\bibinfo {author} {\bibfnamefont {Maxim}\ \bibnamefont
  {Mostovoy}},\ }\bibfield  {title} {\enquote {\bibinfo {title}
  {Ferroelectricity in spiral magnets},}\ }\href {\doibase
  10.1103/PhysRevLett.96.067601} {\bibfield  {journal} {\bibinfo  {journal}
  {Phys. Rev. Lett.}\ }\textbf {\bibinfo {volume} {96}},\ \bibinfo {pages}
  {067601} (\bibinfo {year} {2006})}\BibitemShut {NoStop}%
\bibitem [{\citenamefont {Khomskii}(2009)}]{khomskii_classifying_2009}%
  \BibitemOpen
  \bibfield  {author} {\bibinfo {author} {\bibfnamefont {Daniel}\ \bibnamefont
  {Khomskii}},\ }\bibfield  {title} {\enquote {\bibinfo {title} {Classifying
  multiferroics: {Mechanisms} and effects},}\ }\href {\doibase
  10.1103/Physics.2.20} {\bibfield  {journal} {\bibinfo  {journal} {Physics}\
  }\textbf {\bibinfo {volume} {2}},\ \bibinfo {pages} {20} (\bibinfo {year}
  {2009})}\BibitemShut {NoStop}%
\bibitem [{\citenamefont {Sergienko}\ and\ \citenamefont
  {Dagotto}(2006)}]{PhysRevB.73.094434}%
  \BibitemOpen
  \bibfield  {author} {\bibinfo {author} {\bibfnamefont {I.~A.}\ \bibnamefont
  {Sergienko}}\ and\ \bibinfo {author} {\bibfnamefont {E.}~\bibnamefont
  {Dagotto}},\ }\bibfield  {title} {\enquote {\bibinfo {title} {Role of the
  dzyaloshinskii-moriya interaction in multiferroic perovskites},}\ }\href
  {\doibase 10.1103/PhysRevB.73.094434} {\bibfield  {journal} {\bibinfo
  {journal} {Phys. Rev. B}\ }\textbf {\bibinfo {volume} {73}},\ \bibinfo
  {pages} {094434} (\bibinfo {year} {2006})}\BibitemShut {NoStop}%
\bibitem [{\citenamefont {Kumar}\ \emph {et~al.}(2010)\citenamefont {Kumar},
  \citenamefont {van~den Brink},\ and\ \citenamefont
  {Kampf}}]{PhysRevLett.104.017201}%
  \BibitemOpen
  \bibfield  {author} {\bibinfo {author} {\bibfnamefont {Sanjeev}\ \bibnamefont
  {Kumar}}, \bibinfo {author} {\bibfnamefont {Jeroen}\ \bibnamefont {van~den
  Brink}}, \ and\ \bibinfo {author} {\bibfnamefont {Arno~P.}\ \bibnamefont
  {Kampf}},\ }\bibfield  {title} {\enquote {\bibinfo {title} {Spin-spiral
  states in undoped manganites: Role of finite hund's rule coupling},}\ }\href
  {\doibase 10.1103/PhysRevLett.104.017201} {\bibfield  {journal} {\bibinfo
  {journal} {Phys. Rev. Lett.}\ }\textbf {\bibinfo {volume} {104}},\ \bibinfo
  {pages} {017201} (\bibinfo {year} {2010})}\BibitemShut {NoStop}%
\bibitem [{\citenamefont {Dong}\ \emph {et~al.}(2008)\citenamefont {Dong},
  \citenamefont {Yu}, \citenamefont {Yunoki}, \citenamefont {Liu},\ and\
  \citenamefont {Dagotto}}]{PhysRevB.78.155121}%
  \BibitemOpen
  \bibfield  {author} {\bibinfo {author} {\bibfnamefont {Shuai}\ \bibnamefont
  {Dong}}, \bibinfo {author} {\bibfnamefont {Rong}\ \bibnamefont {Yu}},
  \bibinfo {author} {\bibfnamefont {Seiji}\ \bibnamefont {Yunoki}}, \bibinfo
  {author} {\bibfnamefont {J.-M.}\ \bibnamefont {Liu}}, \ and\ \bibinfo
  {author} {\bibfnamefont {Elbio}\ \bibnamefont {Dagotto}},\ }\bibfield
  {title} {\enquote {\bibinfo {title} {Origin of multiferroic spiral spin order
  in the $r{\text{mno}}_{3}$ perovskites},}\ }\href {\doibase
  10.1103/PhysRevB.78.155121} {\bibfield  {journal} {\bibinfo  {journal} {Phys.
  Rev. B}\ }\textbf {\bibinfo {volume} {78}},\ \bibinfo {pages} {155121}
  (\bibinfo {year} {2008})}\BibitemShut {NoStop}%
\bibitem [{\citenamefont {Azhar}\ and\ \citenamefont
  {Mostovoy}(2017)}]{PhysRevLett.118.027203}%
  \BibitemOpen
  \bibfield  {author} {\bibinfo {author} {\bibfnamefont {Maria}\ \bibnamefont
  {Azhar}}\ and\ \bibinfo {author} {\bibfnamefont {Maxim}\ \bibnamefont
  {Mostovoy}},\ }\bibfield  {title} {\enquote {\bibinfo {title} {Incommensurate
  spiral order from double-exchange interactions},}\ }\href {\doibase
  10.1103/PhysRevLett.118.027203} {\bibfield  {journal} {\bibinfo  {journal}
  {Phys. Rev. Lett.}\ }\textbf {\bibinfo {volume} {118}},\ \bibinfo {pages}
  {027203} (\bibinfo {year} {2017})}\BibitemShut {NoStop}%
\bibitem [{\citenamefont {Banerjee}\ \emph {et~al.}(2014)\citenamefont
  {Banerjee}, \citenamefont {Rowland}, \citenamefont {Erten},\ and\
  \citenamefont {Randeria}}]{PhysRevX.4.031045}%
  \BibitemOpen
  \bibfield  {author} {\bibinfo {author} {\bibfnamefont {Sumilan}\ \bibnamefont
  {Banerjee}}, \bibinfo {author} {\bibfnamefont {James}\ \bibnamefont
  {Rowland}}, \bibinfo {author} {\bibfnamefont {Onur}\ \bibnamefont {Erten}}, \
  and\ \bibinfo {author} {\bibfnamefont {Mohit}\ \bibnamefont {Randeria}},\
  }\bibfield  {title} {\enquote {\bibinfo {title} {Enhanced stability of
  skyrmions in two-dimensional chiral magnets with rashba spin-orbit
  coupling},}\ }\href {\doibase 10.1103/PhysRevX.4.031045} {\bibfield
  {journal} {\bibinfo  {journal} {Phys. Rev. X}\ }\textbf {\bibinfo {volume}
  {4}},\ \bibinfo {pages} {031045} (\bibinfo {year} {2014})}\BibitemShut
  {NoStop}%
\bibitem [{\citenamefont {Yunoki}\ \emph {et~al.}(1998)\citenamefont {Yunoki},
  \citenamefont {Hu}, \citenamefont {Malvezzi}, \citenamefont {Moreo},
  \citenamefont {Furukawa},\ and\ \citenamefont {Dagotto}}]{yunoki_phase_1998}%
  \BibitemOpen
  \bibfield  {author} {\bibinfo {author} {\bibfnamefont {S.}~\bibnamefont
  {Yunoki}}, \bibinfo {author} {\bibfnamefont {J.}~\bibnamefont {Hu}}, \bibinfo
  {author} {\bibfnamefont {A.~L.}\ \bibnamefont {Malvezzi}}, \bibinfo {author}
  {\bibfnamefont {A.}~\bibnamefont {Moreo}}, \bibinfo {author} {\bibfnamefont
  {N.}~\bibnamefont {Furukawa}}, \ and\ \bibinfo {author} {\bibfnamefont
  {E.}~\bibnamefont {Dagotto}},\ }\bibfield  {title} {\enquote {\bibinfo
  {title} {Phase separation in electronic models for manganites},}\ }\href
  {\doibase 10.1103/PhysRevLett.80.845} {\bibfield  {journal} {\bibinfo
  {journal} {Phys. Rev. Lett.}\ }\textbf {\bibinfo {volume} {80}},\ \bibinfo
  {pages} {845--848} (\bibinfo {year} {1998})}\BibitemShut {NoStop}%
\bibitem [{\citenamefont {Gulacsi}(2004)}]{gulacsi_one_2004}%
  \BibitemOpen
  \bibfield  {author} {\bibinfo {author} {\bibfnamefont {M.}~\bibnamefont
  {Gulacsi}},\ }\bibfield  {title} {\enquote {\bibinfo {title} {The one
  dimensional {Kondo} lattice model at partial band filling},}\ }\href
  {\doibase 10.1080/00018730412331313997} {\bibfield  {journal} {\bibinfo
  {journal} {Advances in Physics}\ }\textbf {\bibinfo {volume} {53}},\ \bibinfo
  {pages} {769--937} (\bibinfo {year} {2004})},\ \bibinfo {note} {arXiv:
  cond-mat/0502069}\BibitemShut {NoStop}%
\bibitem [{\citenamefont {Hewson}(1997)}]{HewsonBook}%
  \BibitemOpen
  \bibfield  {author} {\bibinfo {author} {\bibfnamefont {A.~C.}\ \bibnamefont
  {Hewson}},\ }\href@noop {} {\emph {\bibinfo {title} {The Kondo Problem to
  Heavy Fermions}}}\ (\bibinfo  {publisher} {Cambridge University Press},\
  \bibinfo {address} {Cambridge, United Kingdom},\ \bibinfo {year}
  {1997})\BibitemShut {NoStop}%
\bibitem [{\citenamefont {Nagaosa}\ \emph {et~al.}(2010)\citenamefont
  {Nagaosa}, \citenamefont {Sinova}, \citenamefont {Onoda}, \citenamefont
  {MacDonald},\ and\ \citenamefont {Ong}}]{RevModPhys.82.1539}%
  \BibitemOpen
  \bibfield  {author} {\bibinfo {author} {\bibfnamefont {Naoto}\ \bibnamefont
  {Nagaosa}}, \bibinfo {author} {\bibfnamefont {Jairo}\ \bibnamefont {Sinova}},
  \bibinfo {author} {\bibfnamefont {Shigeki}\ \bibnamefont {Onoda}}, \bibinfo
  {author} {\bibfnamefont {A.~H.}\ \bibnamefont {MacDonald}}, \ and\ \bibinfo
  {author} {\bibfnamefont {N.~P.}\ \bibnamefont {Ong}},\ }\bibfield  {title}
  {\enquote {\bibinfo {title} {Anomalous hall effect},}\ }\href {\doibase
  10.1103/RevModPhys.82.1539} {\bibfield  {journal} {\bibinfo  {journal} {Rev.
  Mod. Phys.}\ }\textbf {\bibinfo {volume} {82}},\ \bibinfo {pages}
  {1539--1592} (\bibinfo {year} {2010})}\BibitemShut {NoStop}%
\bibitem [{\citenamefont {Aubry}\ and\ \citenamefont
  {Andr{\'e}}(1980)}]{Aubry_Analyticity_1980}%
  \BibitemOpen
  \bibfield  {author} {\bibinfo {author} {\bibfnamefont {Serge}\ \bibnamefont
  {Aubry}}\ and\ \bibinfo {author} {\bibfnamefont {Gilles}\ \bibnamefont
  {Andr{\'e}}},\ }\bibfield  {title} {\enquote {\bibinfo {title} {Analyticity
  breaking and anderson localization in incommensurate lattices},}\ }\href@noop
  {} {\bibfield  {journal} {\bibinfo  {journal} {Ann. Israel Phys. Soc}\
  }\textbf {\bibinfo {volume} {3}},\ \bibinfo {pages} {18} (\bibinfo {year}
  {1980})}\BibitemShut {NoStop}%
\bibitem [{\citenamefont {Harper}(1955)}]{Harper_Single_1955}%
  \BibitemOpen
  \bibfield  {author} {\bibinfo {author} {\bibfnamefont {Philip~George}\
  \bibnamefont {Harper}},\ }\bibfield  {title} {\enquote {\bibinfo {title}
  {Single band motion of conduction electrons in a uniform magnetic field},}\
  }\href@noop {} {\bibfield  {journal} {\bibinfo  {journal} {Proceedings of the
  Physical Society. Section A}\ }\textbf {\bibinfo {volume} {68}},\ \bibinfo
  {pages} {874} (\bibinfo {year} {1955})}\BibitemShut {NoStop}%
\bibitem [{\citenamefont {Hofstadter}(1976)}]{HofstadterModel}%
  \BibitemOpen
  \bibfield  {author} {\bibinfo {author} {\bibfnamefont {Douglas~R.}\
  \bibnamefont {Hofstadter}},\ }\bibfield  {title} {\enquote {\bibinfo {title}
  {Energy levels and wave functions of bloch electrons in rational and
  irrational magnetic fields},}\ }\href {\doibase 10.1103/PhysRevB.14.2239}
  {\bibfield  {journal} {\bibinfo  {journal} {Phys. Rev. B}\ }\textbf {\bibinfo
  {volume} {14}},\ \bibinfo {pages} {2239--2249} (\bibinfo {year}
  {1976})}\BibitemShut {NoStop}%
\bibitem [{\citenamefont {Laughlin}(1981)}]{PhysRevB.23.5632}%
  \BibitemOpen
  \bibfield  {author} {\bibinfo {author} {\bibfnamefont {R.~B.}\ \bibnamefont
  {Laughlin}},\ }\bibfield  {title} {\enquote {\bibinfo {title} {Quantized hall
  conductivity in two dimensions},}\ }\href {\doibase 10.1103/PhysRevB.23.5632}
  {\bibfield  {journal} {\bibinfo  {journal} {Phys. Rev. B}\ }\textbf {\bibinfo
  {volume} {23}},\ \bibinfo {pages} {5632--5633} (\bibinfo {year}
  {1981})}\BibitemShut {NoStop}%
\bibitem [{\citenamefont {Thouless}\ \emph {et~al.}(1982)\citenamefont
  {Thouless}, \citenamefont {Kohmoto}, \citenamefont {Nightingale},\ and\
  \citenamefont {den Nijs}}]{PhysRevLett.49.405}%
  \BibitemOpen
  \bibfield  {author} {\bibinfo {author} {\bibfnamefont {D.~J.}\ \bibnamefont
  {Thouless}}, \bibinfo {author} {\bibfnamefont {M.}~\bibnamefont {Kohmoto}},
  \bibinfo {author} {\bibfnamefont {M.~P.}\ \bibnamefont {Nightingale}}, \ and\
  \bibinfo {author} {\bibfnamefont {M.}~\bibnamefont {den Nijs}},\ }\bibfield
  {title} {\enquote {\bibinfo {title} {Quantized hall conductance in a
  two-dimensional periodic potential},}\ }\href {\doibase
  10.1103/PhysRevLett.49.405} {\bibfield  {journal} {\bibinfo  {journal} {Phys.
  Rev. Lett.}\ }\textbf {\bibinfo {volume} {49}},\ \bibinfo {pages} {405--408}
  (\bibinfo {year} {1982})}\BibitemShut {NoStop}%
\bibitem [{\citenamefont {Hatsugai}(1993)}]{PhysRevLett.71.3697}%
  \BibitemOpen
  \bibfield  {author} {\bibinfo {author} {\bibfnamefont {Yasuhiro}\
  \bibnamefont {Hatsugai}},\ }\bibfield  {title} {\enquote {\bibinfo {title}
  {Chern number and edge states in the integer quantum hall effect},}\ }\href
  {\doibase 10.1103/PhysRevLett.71.3697} {\bibfield  {journal} {\bibinfo
  {journal} {Phys. Rev. Lett.}\ }\textbf {\bibinfo {volume} {71}},\ \bibinfo
  {pages} {3697--3700} (\bibinfo {year} {1993})}\BibitemShut {NoStop}%
\bibitem [{\citenamefont {Li}\ \emph {et~al.}(2017)\citenamefont {Li},
  \citenamefont {Li},\ and\ \citenamefont {Das~Sarma}}]{PhysRevB.96.085119}%
  \BibitemOpen
  \bibfield  {author} {\bibinfo {author} {\bibfnamefont {Xiao}\ \bibnamefont
  {Li}}, \bibinfo {author} {\bibfnamefont {Xiaopeng}\ \bibnamefont {Li}}, \
  and\ \bibinfo {author} {\bibfnamefont {S.}~\bibnamefont {Das~Sarma}},\
  }\bibfield  {title} {\enquote {\bibinfo {title} {Mobility edges in
  one-dimensional bichromatic incommensurate potentials},}\ }\href {\doibase
  10.1103/PhysRevB.96.085119} {\bibfield  {journal} {\bibinfo  {journal} {Phys.
  Rev. B}\ }\textbf {\bibinfo {volume} {96}},\ \bibinfo {pages} {085119}
  (\bibinfo {year} {2017})}\BibitemShut {NoStop}%
\bibitem [{\citenamefont {Kraus}\ \emph {et~al.}(2012)\citenamefont {Kraus},
  \citenamefont {Lahini}, \citenamefont {Ringel}, \citenamefont {Verbin},\ and\
  \citenamefont {Zilberberg}}]{Kraus_Topological_2012}%
  \BibitemOpen
  \bibfield  {author} {\bibinfo {author} {\bibfnamefont {Yaacov~E.}\
  \bibnamefont {Kraus}}, \bibinfo {author} {\bibfnamefont {Yoav}\ \bibnamefont
  {Lahini}}, \bibinfo {author} {\bibfnamefont {Zohar}\ \bibnamefont {Ringel}},
  \bibinfo {author} {\bibfnamefont {Mor}\ \bibnamefont {Verbin}}, \ and\
  \bibinfo {author} {\bibfnamefont {Oded}\ \bibnamefont {Zilberberg}},\
  }\bibfield  {title} {\enquote {\bibinfo {title} {Topological states and
  adiabatic pumping in quasicrystals},}\ }\href {\doibase
  10.1103/PhysRevLett.109.106402} {\bibfield  {journal} {\bibinfo  {journal}
  {Phys. Rev. Lett.}\ }\textbf {\bibinfo {volume} {109}},\ \bibinfo {pages}
  {106402} (\bibinfo {year} {2012})}\BibitemShut {NoStop}%
\bibitem [{\citenamefont {Avron}\ \emph {et~al.}(1983)\citenamefont {Avron},
  \citenamefont {Seiler},\ and\ \citenamefont {Simon}}]{PhysRevLett.51.51}%
  \BibitemOpen
  \bibfield  {author} {\bibinfo {author} {\bibfnamefont {J.~E.}\ \bibnamefont
  {Avron}}, \bibinfo {author} {\bibfnamefont {R.}~\bibnamefont {Seiler}}, \
  and\ \bibinfo {author} {\bibfnamefont {B.}~\bibnamefont {Simon}},\ }\bibfield
   {title} {\enquote {\bibinfo {title} {Homotopy and quantization in condensed
  matter physics},}\ }\href {\doibase 10.1103/PhysRevLett.51.51} {\bibfield
  {journal} {\bibinfo  {journal} {Phys. Rev. Lett.}\ }\textbf {\bibinfo
  {volume} {51}},\ \bibinfo {pages} {51--53} (\bibinfo {year}
  {1983})}\BibitemShut {NoStop}%
\bibitem [{\citenamefont {Niu}\ and\ \citenamefont
  {Thouless}(1987)}]{PhysRevB.35.2188}%
  \BibitemOpen
  \bibfield  {author} {\bibinfo {author} {\bibfnamefont {Qian}\ \bibnamefont
  {Niu}}\ and\ \bibinfo {author} {\bibfnamefont {D.~J.}\ \bibnamefont
  {Thouless}},\ }\bibfield  {title} {\enquote {\bibinfo {title} {Quantum hall
  effect with realistic boundary conditions},}\ }\href {\doibase
  10.1103/PhysRevB.35.2188} {\bibfield  {journal} {\bibinfo  {journal} {Phys.
  Rev. B}\ }\textbf {\bibinfo {volume} {35}},\ \bibinfo {pages} {2188--2197}
  (\bibinfo {year} {1987})}\BibitemShut {NoStop}%
\bibitem [{\citenamefont {Li}\ \emph {et~al.}(2009)\citenamefont {Li},
  \citenamefont {Chu}, \citenamefont {Jain},\ and\ \citenamefont
  {Shen}}]{PhysRevLett.102.136806}%
  \BibitemOpen
  \bibfield  {author} {\bibinfo {author} {\bibfnamefont {Jian}\ \bibnamefont
  {Li}}, \bibinfo {author} {\bibfnamefont {Rui-Lin}\ \bibnamefont {Chu}},
  \bibinfo {author} {\bibfnamefont {J.~K.}\ \bibnamefont {Jain}}, \ and\
  \bibinfo {author} {\bibfnamefont {Shun-Qing}\ \bibnamefont {Shen}},\
  }\bibfield  {title} {\enquote {\bibinfo {title} {Topological anderson
  insulator},}\ }\href {\doibase 10.1103/PhysRevLett.102.136806} {\bibfield
  {journal} {\bibinfo  {journal} {Phys. Rev. Lett.}\ }\textbf {\bibinfo
  {volume} {102}},\ \bibinfo {pages} {136806} (\bibinfo {year}
  {2009})}\BibitemShut {NoStop}%
\bibitem [{\citenamefont {Groth}\ \emph {et~al.}(2009)\citenamefont {Groth},
  \citenamefont {Wimmer}, \citenamefont {Akhmerov}, \citenamefont
  {Tworzyd\l{}o},\ and\ \citenamefont {Beenakker}}]{PhysRevLett.103.196805}%
  \BibitemOpen
  \bibfield  {author} {\bibinfo {author} {\bibfnamefont {C.~W.}\ \bibnamefont
  {Groth}}, \bibinfo {author} {\bibfnamefont {M.}~\bibnamefont {Wimmer}},
  \bibinfo {author} {\bibfnamefont {A.~R.}\ \bibnamefont {Akhmerov}}, \bibinfo
  {author} {\bibfnamefont {J.}~\bibnamefont {Tworzyd\l{}o}}, \ and\ \bibinfo
  {author} {\bibfnamefont {C.~W.~J.}\ \bibnamefont {Beenakker}},\ }\bibfield
  {title} {\enquote {\bibinfo {title} {Theory of the topological anderson
  insulator},}\ }\href {\doibase 10.1103/PhysRevLett.103.196805} {\bibfield
  {journal} {\bibinfo  {journal} {Phys. Rev. Lett.}\ }\textbf {\bibinfo
  {volume} {103}},\ \bibinfo {pages} {196805} (\bibinfo {year}
  {2009})}\BibitemShut {NoStop}%
\bibitem [{\citenamefont {Su}\ \emph {et~al.}(2016)\citenamefont {Su},
  \citenamefont {Avishai},\ and\ \citenamefont {Wang}}]{PhysRevB.93.214206}%
  \BibitemOpen
  \bibfield  {author} {\bibinfo {author} {\bibfnamefont {Ying}\ \bibnamefont
  {Su}}, \bibinfo {author} {\bibfnamefont {Y.}~\bibnamefont {Avishai}}, \ and\
  \bibinfo {author} {\bibfnamefont {X.~R.}\ \bibnamefont {Wang}},\ }\bibfield
  {title} {\enquote {\bibinfo {title} {Topological anderson insulators in
  systems without time-reversal symmetry},}\ }\href {\doibase
  10.1103/PhysRevB.93.214206} {\bibfield  {journal} {\bibinfo  {journal} {Phys.
  Rev. B}\ }\textbf {\bibinfo {volume} {93}},\ \bibinfo {pages} {214206}
  (\bibinfo {year} {2016})}\BibitemShut {NoStop}%
\bibitem [{\citenamefont {Bak}\ and\ \citenamefont
  {Jensen}(1980)}]{bak_theory_1980}%
  \BibitemOpen
  \bibfield  {author} {\bibinfo {author} {\bibfnamefont {P.}~\bibnamefont
  {Bak}}\ and\ \bibinfo {author} {\bibfnamefont {M.~H.}\ \bibnamefont
  {Jensen}},\ }\bibfield  {title} {\enquote {\bibinfo {title} {Theory of
  helical magnetic structures and phase transitions in {MnSi} and {FeGe}},}\
  }\href {\doibase 10.1088/0022-3719/13/31/002} {\bibfield  {journal} {\bibinfo
   {journal} {J. Phys. C: Solid State Phys.}\ }\textbf {\bibinfo {volume}
  {13}},\ \bibinfo {pages} {L881} (\bibinfo {year} {1980})}\BibitemShut
  {NoStop}%
\bibitem [{\citenamefont {Resta}(1994)}]{RevModPhys.66.899}%
  \BibitemOpen
  \bibfield  {author} {\bibinfo {author} {\bibfnamefont {Raffaele}\
  \bibnamefont {Resta}},\ }\bibfield  {title} {\enquote {\bibinfo {title}
  {Macroscopic polarization in crystalline dielectrics: the geometric phase
  approach},}\ }\href {\doibase 10.1103/RevModPhys.66.899} {\bibfield
  {journal} {\bibinfo  {journal} {Rev. Mod. Phys.}\ }\textbf {\bibinfo {volume}
  {66}},\ \bibinfo {pages} {899--915} (\bibinfo {year} {1994})}\BibitemShut
  {NoStop}%
\bibitem [{\citenamefont {Thouless}(1983)}]{PhysRevB.27.6083}%
  \BibitemOpen
  \bibfield  {author} {\bibinfo {author} {\bibfnamefont {D.~J.}\ \bibnamefont
  {Thouless}},\ }\bibfield  {title} {\enquote {\bibinfo {title} {Quantization
  of particle transport},}\ }\href {\doibase 10.1103/PhysRevB.27.6083}
  {\bibfield  {journal} {\bibinfo  {journal} {Phys. Rev. B}\ }\textbf {\bibinfo
  {volume} {27}},\ \bibinfo {pages} {6083--6087} (\bibinfo {year}
  {1983})}\BibitemShut {NoStop}%
\bibitem [{\citenamefont {Yu}\ \emph {et~al.}(2011)\citenamefont {Yu},
  \citenamefont {Qi}, \citenamefont {Bernevig}, \citenamefont {Fang},\ and\
  \citenamefont {Dai}}]{PhysRevB.84.075119}%
  \BibitemOpen
  \bibfield  {author} {\bibinfo {author} {\bibfnamefont {Rui}\ \bibnamefont
  {Yu}}, \bibinfo {author} {\bibfnamefont {Xiao~Liang}\ \bibnamefont {Qi}},
  \bibinfo {author} {\bibfnamefont {Andrei}\ \bibnamefont {Bernevig}}, \bibinfo
  {author} {\bibfnamefont {Zhong}\ \bibnamefont {Fang}}, \ and\ \bibinfo
  {author} {\bibfnamefont {Xi}~\bibnamefont {Dai}},\ }\bibfield  {title}
  {\enquote {\bibinfo {title} {Equivalent expression of ${\mathbb{z}}_{2}$
  topological invariant for band insulators using the non-abelian berry
  connection},}\ }\href {\doibase 10.1103/PhysRevB.84.075119} {\bibfield
  {journal} {\bibinfo  {journal} {Phys. Rev. B}\ }\textbf {\bibinfo {volume}
  {84}},\ \bibinfo {pages} {075119} (\bibinfo {year} {2011})}\BibitemShut
  {NoStop}%
\bibitem [{\citenamefont {Asb\'{o}th}\ \emph {et~al.}(2016)\citenamefont
  {Asb\'{o}th}, \citenamefont {Oroszl\'{a}ny},\ and\ \citenamefont
  {P\'{a}lyi}}]{TopoBook}%
  \BibitemOpen
  \bibfield  {author} {\bibinfo {author} {\bibfnamefont {J.~K.}\ \bibnamefont
  {Asb\'{o}th}}, \bibinfo {author} {\bibfnamefont {L.}~\bibnamefont
  {Oroszl\'{a}ny}}, \ and\ \bibinfo {author} {\bibfnamefont {A.}~\bibnamefont
  {P\'{a}lyi}},\ }\href@noop {} {\emph {\bibinfo {title} {A Short Course on
  Topological Insulators}}}\ (\bibinfo  {publisher} {Springer},\ \bibinfo
  {address} {New York, USA},\ \bibinfo {year} {2016})\BibitemShut {NoStop}%
\bibitem [{\citenamefont {Fukui}\ \emph {et~al.}(2005)\citenamefont {Fukui},
  \citenamefont {Hatsugai},\ and\ \citenamefont {Suzuki}}]{fukui_chern_2005}%
  \BibitemOpen
  \bibfield  {author} {\bibinfo {author} {\bibfnamefont {Takahiro}\
  \bibnamefont {Fukui}}, \bibinfo {author} {\bibfnamefont {Yasuhiro}\
  \bibnamefont {Hatsugai}}, \ and\ \bibinfo {author} {\bibfnamefont {Hiroshi}\
  \bibnamefont {Suzuki}},\ }\bibfield  {title} {\enquote {\bibinfo {title}
  {Chern numbers in discretized brillouin zone: efficient method of computing
  (spin) hall conductances},}\ }\href {https://doi.org/10.1143/JPSJ.74.1674}
  {\bibfield  {journal} {\bibinfo  {journal} {Journal of the Physical Society
  of Japan}\ }\textbf {\bibinfo {volume} {74}},\ \bibinfo {pages} {1674--1677}
  (\bibinfo {year} {2005})}\BibitemShut {NoStop}%
\end{thebibliography}%

\end{document}